\documentclass[
 superscriptaddress,
 twocolumn,
 amsmath,amssymb,
 aps,
]{revtex4-2}

\usepackage{graphicx}
\usepackage{dcolumn}
\usepackage{bm}
\usepackage{enumerate}
\usepackage{threeparttable}
\usepackage{multirow}
\usepackage{makecell}

\begin{document}


\title{Hyperspin Altermagnets}

\author{Hai-Yang Ma}
\thanks{Corresponding author: mahaiyang@quantumsc.cn}
\affiliation{Quantum Science Center of Guangdong-Hong Kong-Macao Greater Bay Area, Shenzhen 518045, China}

\author{Yuanchang Li}
\thanks{Corresponding author: yuancli@bit.edu.cn}
\affiliation{School of Interdisciplinary Science, Beijing Institute of Technology, Beijing 100081, China}
\affiliation{Key Lab of Advanced Optoelectronic Quantum Architecture and Measurement (MOE), and School of Physics, Beijing Institute of Technology, Beijing 100081, China}

\author{Hu Xu}
\affiliation{Department of Physics, Southern University of Science and Technology, Shenzhen 518055, China}

\author{Shengbai Zhang}
\thanks{Corresponding author: zhangs9@rpi.edu}
\affiliation{Department of Physics, Applied Physics and Astronomy, Rensselaer Polytechnic Institute, Troy, New York 12180, USA}

\author{Jin-Feng Jia}
\thanks{Corresponding author: jfjia@sjtu.edu.cn}
\affiliation{Quantum Science Center of Guangdong-Hong Kong-Macao Greater Bay Area, Shenzhen 518045, China}
\affiliation{Department of Physics, Southern University of Science and Technology, Shenzhen 518055, China}
\affiliation{Key Laboratory of Artificial Structures and Quantum Control (Ministry of Education), TD Lee institute, School of Physics and Astronomy, Shanghai Jiao Tong University, 800 Dongchuan Road, Shanghai 200240, China}
\affiliation{Hefei National Laboratory, Hefei 230088, China}


\begin{abstract}
The behavior of spin quantum in $\mathbf{k}$-space is key to identifying altermagnets (AMs) as the third kind of fundamental collinear magnetism. In contrast, non-collinear magnets—though abundant in nature—lack well-defined spin quantum numbers, and the resulting spin textures are often highly complex, which limits their potential for next-generation spintronic applications. Here we propose hyperspin, which lives in a higher-dimensional space, to address these drawbacks. Through analyzing the commutation relations between spin and Hamiltonian for a class of non-collinear magnets, we reveal it is a hyperspin, rather than the usual spin, that commutes with Hamiltonian. Unexpectedly, these \textit{non-collinear} magnets should also show \textit{collinear} spin-split bands in $\mathbf{k}$-space like collinear AMs. We therefore classify such non-collinear magnets as hyperspin altermagnets (HAMs), as opposed to the usual collinear AMs. Our theory elucidates the fundamental physics of AMs and HAMs and provides a framework for exploring the wide range of non-collinear magnets that may possess other kinds of conserved quantities.  
\end{abstract}

\keywords{ }
\maketitle


\section[\label{sec.1}]{Introduction}
\begin{figure*}[tb]
\includegraphics[width=1.0\textwidth]{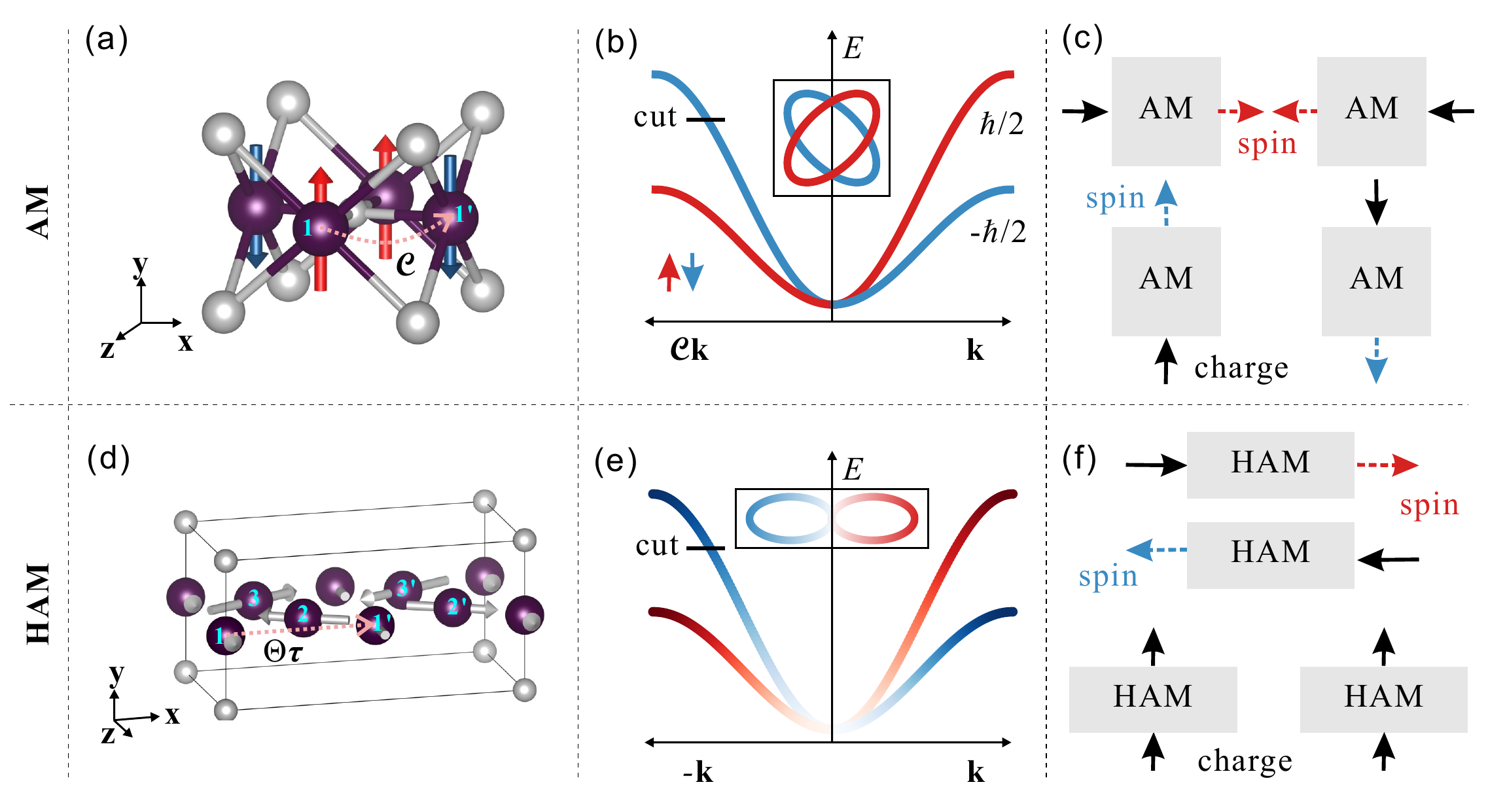}
\caption{\label{fig1} (a-c) AM versus (d-f) HAM. (a) AMs are collinear in real space, and the two magnetic atoms 1 and 1' are related by a crystal symmetry $\mathcal{C}$. (d) HAMs are coplanar but non-collinear in real space, the first set of three magnetic atoms 1, 2, 3 is related to the second set of three magnetic atoms 1$'$, 2$'$, 3$'$ by a $\Theta\bm{\tau}$ symmetry. (b) AMs have alternating spin-split bands, whose spin $\langle\hat{\mathcal{S}}_y\rangle$ is quantized and the spin-up band at $\mathbf{k}$ is degenerate with the spin-down band at $\mathcal{C}\mathbf{k}$. (e) HAMs also have alternating spin-split bands, but different from (b), its out-of-plane spin, say $\langle\hat{\mathcal{S}}_y\rangle\in[-\hbar/2,\ \hbar/2]$, is not quantized and the $\Theta\bm{\tau}$ symmetry requires that the “spin-up” band at $\mathbf{k}$ is degenerate with the “spin-down” band at -$\mathbf{k}$. Insets in (b) and (e) are typical constant energy contours. (c) and (f), Spin current generation. Details are discussed in the main text.}
\end{figure*}

The emergence of altermagnets (AMs) \cite{vsmejkal2022emerging,PhysRevX.12.031042,mazin2022altermagnetism}, or crystalline symmetry ($\mathcal{C}$)-paired spin-valley locking materials \cite{ma2021multifunctional}, has broken the traditional dichotomy between ferromagnets and antiferromagnets, establishing a third class of magnetic order. In real space, like conventional antiferromagnets, AMs have zero macroscopic magnetization [Fig.~\ref{fig1}(a)]; while in $\mathbf{k}$-space, like ferromagnets, AMs have collinear spin-split bands, but the spin splitting alternates as $E_{n\uparrow}\left(\mathbf{k}\right)=E_{n\downarrow}\left(\mathcal{C}\mathbf{k}\right)$, where $\mathbf{k}$ is the crystal momentum, see Fig.~\ref{fig1}(b). Therefore, AMs bear both the advantages of antiferromagnets, e.g., ultrafast spin dynamics and being inert to external magnetic field \cite{baltz2018antiferromagnetic}, and ferromagnets, e.g., spin current generation \cite{ma2021multifunctional}, as shown in Fig.~\ref{fig1}(c). These attribute AMs with tremendous potential for next-generation spintronic applications \cite{PhysRevX.12.031042,ma2021multifunctional,baltz2018antiferromagnetic}. 

Due to the collinear spin-split bands in k-space, i.e., $\uparrow$ vs. $\downarrow$ for electrons indexed by $\left(n,\mathbf{k}\right)$, it is not surprising that most research on AMs has focused on real-space collinear magnets \cite{krempasky2024altermagnetic,amin2024nanoscale,ding2024large,jiang2025metallic,zhang2025crystal,ma2024altermagnetic}, where spin is a good quantum number. Despite the abundance of non-collinear magnets in nature \cite{zhu2024observation,ma2026multicolor}, see for example Fig.~\ref{fig1}(d), whose intrinsic spin-split bands typically exhibit non-collinear spin textures. However, recent density functional theory (DFT) calculations revealed that some non-collinear magnets with combined time-reversal ($\Theta$) and half translational symmetry ($\bm{\tau}$) exhibit a characteristic alternated collinear spin-split bands typical of the AMs \cite{hellenes2023p}. Furthermore, the odd parity of the spin splitting in such system is complementary to the even parity often found in collinear AMs \cite{hellenes2023p,song2025unified,brekke2024minimal,sukhachov2024impurity,ezawa2024topological,maeda2025classification,chakraborty2025highly,soori2025crossed,nagae2025flat,sukhachov2025coexistence,yamada2025gapping,song2025electrical,zhou2025anisotropic,yamada2025metallic,ezawa2025third,zhu2025floquet}. In the absence of spin-orbit-coupling (SOC), the breaking of both $P\Theta$ ($P$:\ spatial inversion) and $\Theta\bm{\tau}$ symmetries is required for collinear AMs to lift the spin degeneracy. Additional symmetry is also required to ensure the alternated spin splitting, which is typically a rotation or mirror symmetry, but is never $\Theta\bm{\tau}$. While for non-collinear magnets, no such good quantum number exists. These observations challenge our usual understandings of spin: since spin conservation has been violated by the non-collinearity, how could spin still be collinear while non-quantized in k-space? More broadly, what are the physics mechanisms and symmetries that give rise to the collinear and alternating spin-split bands in non-collinear magnets?

In this work, we propose an emergent hyperspin to \textit{analytically} address these issues. We build a general Hamiltonian $\hat{H}$ applicable to a class of non-collinear magnets with dual sublattices interconnected by $\Theta\bm{\tau}$ symmetry, see Fig.~\ref{fig1}(d). Analytical solutions of this Hamiltonian yield hyperspin (defined below) splitting energy bands, which appears to resemble those of collinear AMs, but is governed by a completely different commutation relation. Properties such as odd/even-parity splitting and symmetry protection are direct manifestations of these differing commutation relations. Given the stark distinctions, we classify these real-space non-collinear and coplanar magnets, including the p-wave magnets \cite{hellenes2023p}, as hyperspin altermagnets (HAMs), as opposed to the usual AMs which are collinear in real space.

\section[\label{sec.2}]{AM vs. HAM}
Alternating collinear spin splitting is the fingerprint of altermagnets. Generally, three conditions must be met to achieve this: 1) alternately splitting the spin degeneracy; 2) the presence of a good spin quantum number to ensure that the spin splitting is collinear in the usual way; and 3) an additional symmetry to guarantee complete spin compensation. The good spin quantum number implies the existence of a conserved quantity $\hat{S}$, such that $[\hat{H},\ \hat{S}]=0$. Since $\hat{H}$ already incorporates the system’s symmetry, the commutation relation therefore uniquely characterizes the physical essence of an altermagnet. For AMs, $\hat{S}={\hat{\mathcal{S}}}_y$ \cite{note1}, which is the physical two-component spin angular momentum along the collinear direction [see Fig.~\ref{fig1}(a)]. It is straightforward to show $[\hat{H},\ {\hat{\mathcal{S}}}_y]=0$ and $\langle\hat{\mathcal{S}}_y\rangle$ is quantized as $\pm\hbar/2$, which guarantees both the real-space magnetic order and $\mathbf{k}$-space collinear energy bands. In the presence of a desired symmetry that leads to full spin compensation, collinear and alternating spin-split bands will result \cite{vsmejkal2022emerging,PhysRevX.12.031042,ma2021multifunctional}. For HAMs, the physical spin ${\hat{\mathcal{S}}}_y$ does not commute with $\hat{H}$. However, there exists a hyperspin ${\hat{\mathcal{J}}}_y$, where $y$ is in the direction perpendicular to all local magnetizations, see Fig.~\ref{fig1}(b), such that $[\hat{H},\ {\hat{\mathcal{J}}}_y]=0$. It gives rise to $\langle\hat{\mathcal{J}}_x\rangle=\langle\hat{\mathcal{J}}_z\rangle=\langle\hat{\mathcal{S}}_x\rangle=\langle\hat{\mathcal{S}}_z\rangle=0$, see Sec. I of Ref.~\cite{SM}, which ensures collinear bands in $\mathbf{k}$-space, despite that $\langle\hat{\mathcal{S}}_y\rangle$ is no longer $\pm\hbar/2$. This is precisely the microscopic underpinning that distinguishes real-space non-collinear HAMs from collinear AMs. Furthermore, we show that $[\hat{H},\ {\hat{\mathcal{J}}}_y]=0$ relies on the non-collinear spin order to be coplanar, and, as far as we can tell, $\Theta\bm{\tau}$ symmetry, which is excluded for AMs, is a prerequisite for realizing such HAMs. Due to $\Theta\bm{\tau}$, HAMs must exhibit odd-wave spin splitting [e.g., see inset of Fig.~\ref{fig1}(e)].

\begin{table}
\renewcommand{\arraystretch}{1.2}
\centering
\caption{Comparison of AFMs, AMs and HAMs. For AFM, either $P\Theta$ or $\Theta\bm{\tau}$ is enough to guarantee spin degeneracy, while $P\Theta$ is excluded both in AMs and HAMs. Phrase "$\mathbf{r}$-collinear" denotes that the magnetic order in real space is collinear, whereas “$\mathbf{k}$-collinear” denotes the spin of the bands in $\mathbf{k}$-space is collinear. $\surd$ = true and $\times$ = false, “odd” means the parity of the spin splitting is odd-wave. In all cases, $\langle\hat{\mathcal{S}}_x\rangle=\langle\hat{\mathcal{S}}_z\rangle=0$.}
\label{table1}
\begin{ruledtabular}
    \begin{tabular}{c|cccccc}
      & $P\Theta$ & $\Theta\bm{\tau}$  &  $\mathbf{r}$-collinear & $\mathbf{k}$-collinear & \makecell[c]{spin\\splitting} & \makecell[c]{$\langle\hat{\mathcal{S}}_y\rangle$\\quantized}\\
      \hline
      AFM & $\surd$ & $\surd$ & $\surd$ & $\surd$ & $\times$ & $\surd$\\
      AM & $\times$ & $\times$ & $\surd$ & $\surd$ & $\surd$ & $\surd$\\
      HAM & $\times$ & $\surd$ & $\times$ & $\surd$ & odd & $\times$\\
    \end{tabular}
\end{ruledtabular}
\end{table}

Table~\ref{table1} summarizes major differences between AMs and HAMs, as well as with conventional collinear antiferromagnets (AFM). In terms of applications, both AMs and HAMs can generate spin current \cite{ma2021multifunctional,chakraborty2025highly}, but the effects of even and odd waves are strikingly different. For AMs, as illustrated by d-wave in Fig.~\ref{fig1}(c), when a charge current flows in from the left, a spin-up current flows out from the right. If the direction of the charge current is reversed, the output remains spin-up. To obtain a spin-down current, the input charge current must be switched to a perpendicular direction \cite{ma2021multifunctional}. HAMs behave differently, as illustrated by the p-wave in Fig.~\ref{fig1}(f). When a charge current flows in from the left, a spin-up current flows out from the right. If the direction of the charge current is reversed so that it flows in from the right, what flows out from the left becomes a spin-down current \cite{chakraborty2025highly}. In the vertical direction, however, the charge current does not result in any spin current. This exotic magnetic diode effect resembles the superconducting diode effect \cite{jiang2022superconducting}.

\section[\label{sec.3}]{Spin degeneracy for AFM with $\mathbf{\Theta}\bm{\tau}$}
\begin{figure}[tb]
\includegraphics[width=0.4\textwidth]{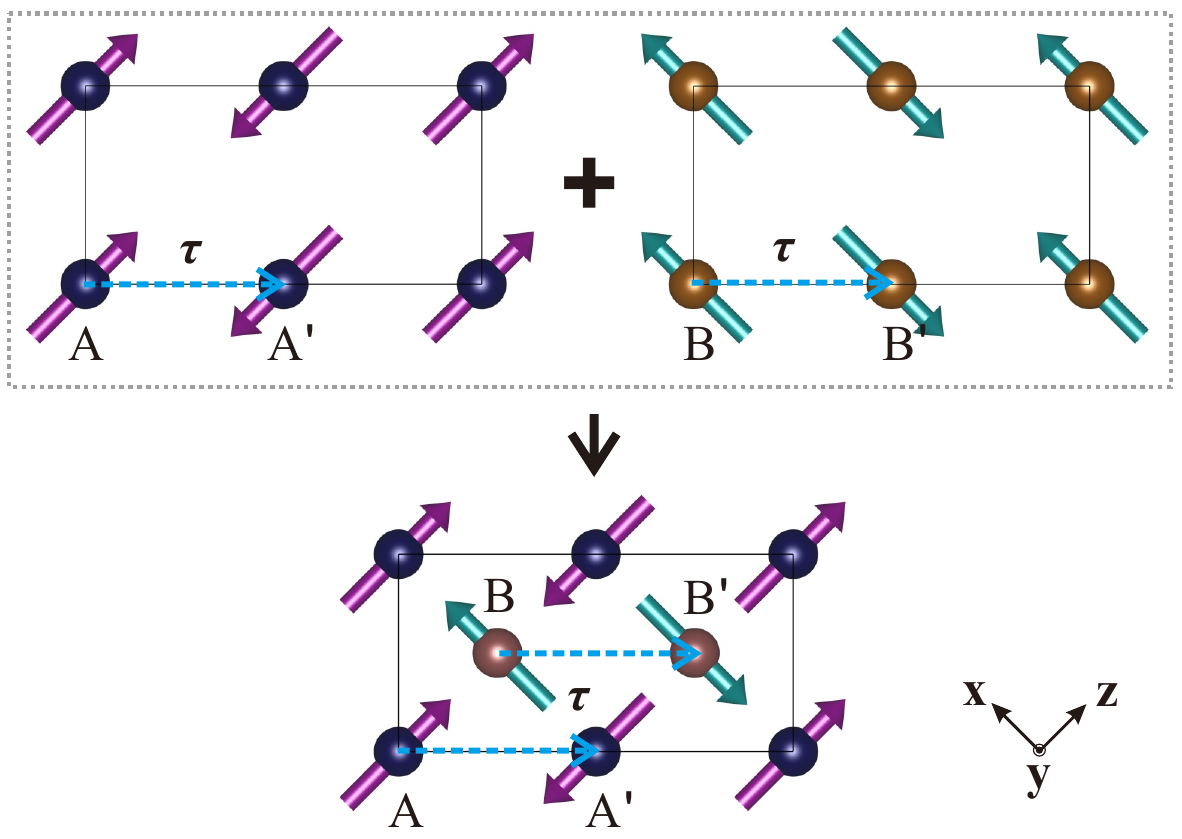}
\caption{\label{fig2} Illustration of how two AFMs are merged into a non-collinear HAM. An AFM lattice contains two magnetic atoms, AA$'$ or BB$'$ in the plots. Arrows indicate the direction of local magnetic moments. The two magnetic atoms A and A$'$ (B and B$'$) within an AFM sublattice are related by $\Theta\bm{\tau}$. A HAM can be built by putting two or more AFM lattices, termed subsystems now, into a single lattice, while keeping the $\Theta\bm{\tau}$ symmetry.}
\end{figure}

Before discussing HAMs, as a prelude, we would like to discuss the effects of $\Theta\bm{\tau}$ on collinear magnets. It is known that either $P\Theta$ or $\Theta\bm{\tau}$ would cause spin degeneracy, $E_{n\uparrow}\left(\mathbf{k}\right)=E_{n\downarrow}\left(\mathbf{k}\right)$, in AFM. For $P\Theta$ symmetry, the spin degeneracy is easily understood. But for $\Theta\bm{\tau}$, this is less obvious, as it appears that $\Theta\bm{\tau}$ can only yield $E_{n\uparrow}\left(\mathbf{k}\right)=E_{n\downarrow}\left(-\mathbf{k}\right)$. Nevertheless, we prove in Sec. IIA-C of Ref.~\cite{SM} that the spin degeneracy indeed holds for $\Theta\bm{\tau}$. Here, we only sketch some of the key steps in such a derivation.
An AFM unit cell has two magnetic atoms A and A$'$ with collinear local moments, $\mathbf{M}_{\rm A}$ and $-\mathbf{M}_{\rm A}$, respectively (see Fig.~\ref{fig2}). Within the single-particle approximation, the Hamiltonian, is given as $\hat{H}\left(\mathbf{r}\right)={\hat{\mathbf{p}}}^2/2m+\sum_{\pm}{V^\pm(\mathbf{r})}$. Using either a plane-wave or Wannier basis, and in combination with the hermiticity of $\hat{H}\left(\mathbf{r}\right)$, one can write in the $\mathbf{k}$-space the diagonalized Hamiltonian under $\Theta\bm{\tau}$ symmetry as (see Sec. IIA-C of Ref.~\cite{SM})
\begin{equation}
\label{ham1}
H_{m,\mathbf{k}}=
    \begin{pmatrix}
    E_{m,\mathbf{k}} & 0\\
    0 & E_{m,\mathbf{k}}
    \end{pmatrix}
\end{equation}
where $E_{m,\mathbf{k}}$ is the eigen energy with $m$ the band index. We will denote the eigenvectors as $|u_{m\mathbf{k}}^+(\mathbf{r})\rangle=|u_{m\mathbf{k}}(\mathbf{r})\rangle\xi^+$ and $|u_{m\mathbf{k}}^-(\mathbf{r})\rangle=|u_{m\mathbf{k}}(\mathbf{r})\rangle\xi^-$, where $|u_{m\mathbf{k}}(\mathbf{r})\rangle$ is the spatial part and $\xi^\pm$ are the spin parts along $\pm\mathbf{M}_{\rm A}$. Under such a basis, the spin operator $\hat{\mathcal{S}}$ has the usual 2$\times$2 Pauli matrix form $\hbar/2\hat{\bm{\sigma}}$.

\section[\label{sec.4}]{Hyperspin operator and the Hamiltonian}
Now we build our HAM system out of the AFM subsystems with $\Theta\bm{\tau}$ symmetry. The simplest one is to merge two AFM subsystems, $\mathbb{A}={\rm AA'}$ and $\mathbb{B}={\rm BB’}$, together with $\mathbf{M}_\mathbb{A}=\mathbf{M}_{\rm A}$ and $\mathbf{M}_\mathbb{B}=\mathbf{M}_{\rm B}$, as shown in Fig.~\ref{fig2}. Before merging, we will take the spin quantization direction, i.e., the $\mathbf{z}$ axis, to be $\mathbf{M}_\mathbb{A}$ for subsystem $\mathbb{A}$ and $\mathbf{M}_\mathbb{B}$ for subsystem $\mathbb{B}$. After merging, we choose the common $\mathbf{z}$ axis along $\mathbf{M}_\mathbb{A}$, the common $\mathbf{y}$ axis along the direction of the cross product, $\mathbf{M}_\mathbb{A}\times\mathbf{M}_\mathbb{B}$, and the common $\mathbf{x}$ axis along $\mathbf{y}\times\mathbf{z}$ (see Fig.~\ref{fig2}). This choice of coordinate is to compromise with $\Theta$ such that a common form of $\Theta=-i\sigma_y\hat{K}$ can be used, where $\hat{K}$ is the complex conjugation. We denote the eigenvectors of subsystems $\mathbb{A}$ and $\mathbb{B}$ as $|u_{\mathbb{A},m\mathbf{k}}^\pm(\mathbf{r})\rangle=|u_{\mathbb{A},m\mathbf{k}}(\mathbf{r})\rangle\xi_\mathbb{A}^\pm$ and $|u_{\mathbb{B},m\mathbf{k}}^\pm(\mathbf{r})\rangle=|u_{\mathbb{B},m\mathbf{k}}(\mathbf{r})\rangle\xi_\mathbb{B}^\pm$, where $\xi_{\mathbb{A}(\mathbb{B})}^\pm$ are local spin states along $\mathbf{M}_\mathbb{A}$ ($\mathbf{M}_\mathbb{B}$), respectively. In the following, they will be used as the basis in the order: $|u_{\mathbb{A},m\mathbf{k}}^+(\mathbf{r})\rangle,|u_{\mathbb{A},m\mathbf{k}}^-(\mathbf{r})\rangle,|u_{\mathbb{B},m\mathbf{k}}^+(\mathbf{r})\rangle,|u_{\mathbb{B},m\mathbf{k}}^-(\mathbf{r})\rangle$, to solve $\hat{H}$. Note that here the choice of exact subsystems is not critical, so long as each preserves the $\Theta\bm{\tau}$ symmetry.

The physical spin is given as $\hat{\mathcal{S}}=\sum_{\mu\mu\prime}{{\hat{O}}_\mu^\dag\hat{\mathcal{S}}{\hat{O}}_{\mu\prime}}$ where ${\hat{O}}_\mu$ is the projection operator \cite{soriano2014theory} that projects into the subspace spanned by the $\mu$th subsystem, $\mu=\mathbb{A},\mathbb{B},\mathbb{C}\cdots$. For our two-subsystem case, it can be represented by
\begin{equation}
\label{spin}
\hat{\mathcal{S}}\doteq\frac{\hbar}{2}
    \begin{pmatrix}
        \hat{\bm{\sigma}} & \hat{\bm{\sigma}}\\
        \hat{\bm{\sigma}} & \hat{\bm{\sigma}}
    \end{pmatrix}.
\end{equation}
Because the basis is not orthogonal, i.e., $\mathrm{\Lambda}_m(\mathbf{k})=\langle u_{\mathbb{A},m\mathbf{k}}(\mathbf{r})| u_{\mathbb{B},m\mathbf{k}}(\mathbf{r})\rangle\neq0$, the off-diagonal blocks of Eq.~(\ref{spin}), $\langle u_{\mathbb{A},m\mathbf{k}}^\pm(\mathbf{r})|\hat{S}| u_{\mathbb{B},m\mathbf{k}}^\pm(\mathbf{r})\rangle$, may not vanish. Even so, we notice that there exists a different operator, to be named here hyperspin, $\hat{\mathcal{J}}=\sum_{\mu}{{\hat{O}}_\mu^\dag\hat{\mathcal{S}}{\hat{O}}_\mu}$, whose $y$-component commutes with $\hat{H}$, i.e., $[\hat{H},\ {\hat{\mathcal{J}}}_y]=0$. More specifically, $\hat{\mathcal{J}}$ is a direct sum of $\hat{\mathcal{S}}$ for subsystems, $\hat{\mathcal{J}}={\hat{\mathcal{S}}}_\mathbb{A}\oplus{\hat{\mathcal{S}}}_\mathbb{B}$, which can be represented by
\begin{equation}
\hat{\mathcal{J}}\doteq\frac{\hbar}{2}
    \begin{pmatrix}
        \hat{\bm{\sigma}} & 0\\
        0 & \hat{\bm{\sigma}}
    \end{pmatrix}
\end{equation}
whose y-component is (details can be found in Sec. III of Ref.~\cite{SM})
\begin{equation}
\hat{\mathcal{J}}_y=\frac{\hbar}{2}
    \begin{pmatrix}
        \sigma_y & 0\\
        0 & \sigma_y
    \end{pmatrix}=\frac{\hbar}{2}
    \begin{pmatrix}
        0 & -i & 0 & 0\\
        i & 0 & 0 & 0\\
        0 & 0 & 0 & -i\\
        0 & 0 & i & 0
    \end{pmatrix},
\end{equation}
with the following double-degenerate eigenvalues and (unnormalized) eigenvectors
\begin{equation}
\label{vec1}
    \begin{aligned}
    j_y&=+\hbar/2:\ \mathbf{w}_1=[-i,1,0,0]^T,\ \mathbf{w}_2=\left[0,0,-i,1\right]^T;\\
    j_y&=-\hbar/2:\ \mathbf{w}_3=[i,1,0,0]^T,\ \mathbf{w}_4=[0,0,i,1]^T.
    \end{aligned}
\end{equation}
In analogy with Eq.~(\ref{ham1}), the $\mathbf{k}$-space Hamiltonian for our HAM system can be written as a 4$\times$4 matrix, indexed, however, by two integers, $m$ and $m'$. Imposing the $\Theta\bm{\tau}$ symmetry, we can write down the diagonal $m=m'$ matrix as
\begin{equation}
\label{ham2}
\begin{aligned}
    H_{mn}(\mathbf{k})=&h_{0,m}(\mathbf{k})I_{4\times4}+\\&
    \begin{bmatrix}
        h_{1,m}(\mathbf{k}) & -ih_{4,m}(\mathbf{k}) & h_{2,m}(\mathbf{k}) & -ih_{3,m}(\mathbf{k})\\
        ih_{4,m}(\mathbf{k}) & h_{1,m}(\mathbf{k}) & ih_{3,m}(\mathbf{k}) & h_{2,m}(\mathbf{k})\\
        h^*_{2,m}(\mathbf{k}) & -ih^*_{3,m}(\mathbf{k}) & -h_{1,m}(\mathbf{k}) & -ih_{5,m}(\mathbf{k})\\
        ih^*_{3,m}(\mathbf{k}) & h^*_{2,m}(\mathbf{k}) & ih_{5,m}(\mathbf{k}) & -h_{1,m}(\mathbf{k})
    \end{bmatrix}
\end{aligned},
\end{equation}
where hermiticity require $h_{0,m}(\mathbf{k})$, $h_{1,m}(\mathbf{k})$, $h_{4,m}(\mathbf{k})$ and $h_{5,m}(\mathbf{k})$ to be real. The physical meaning of the various terms in Eq.~(\ref{ham2}) can be understood using Fig.~\ref{fig2}: $h_{0,m}(\mathbf{k})$ and $h_{1,m}(\mathbf{k})$ are the diagonal terms in Eq.~(\ref{ham1}), $h_{2,m}(\mathbf{k})$ is the inter-subsystem spin-(++) coupling, $h_{3,m}(\mathbf{k})$ is the inter-subsystem spin-($+-$) coupling, $h_{4,m}(\mathbf{k})$ and $h_{5,m}(\mathbf{k})$ are the intra-subsystem spin-($+-$) couplings. Note that, in a broader sense, Eq.~(\ref{ham2}) also applies to collinear AMs for which $h_{3,m}(\mathbf{k})$, $h_{4,m}(\mathbf{k})$ and $h_{5,m}(\mathbf{k})$ vanish, but the inter-subsystem spin-(++) and spin-($--$) couplings, i.e., the two $h_{2,m}(\mathbf{k})$, are no longer the same. Of course, in this case an additional crystal symmetry is required to achieve alternating collinear spin-split bands. Discussions on the off-diagonal $m\neq m'$ Hamiltonian matrix can be found in Sec. IV of Ref.~\cite{SM}.
Due to the commutation relation, simultaneous eigenvectors exist for $\hat{H}$ and ${\hat{\mathcal{J}}}_y$. By taking a linear combination of the eigenvectors in Eq.~(\ref{vec1}), 
\begin{equation}
    \begin{aligned}
        \mathbf{v}_1=s(\mathbf{k})\mathbf{w}_1+e^{-i\phi(\mathbf{k})}\mathbf{w}_2,\ \mathbf{v}_2=-\frac{1}{s(\mathbf{k})}\mathbf{w}_1+e^{-i\phi(\mathbf{k})}\mathbf{w}_2\\
        \mathbf{v}_3=t(\mathbf{k})\mathbf{w}_3+e^{-i\psi(\mathbf{k})}\mathbf{w}_4,\ \mathbf{v}_4=-\frac{1}{t(\mathbf{k})}\mathbf{w}_3+e^{-i\psi(\mathbf{k})}\mathbf{w}_4
    \end{aligned}
\end{equation}
where $s(\mathbf{k})$, $t(\mathbf{k})$, $\phi(\mathbf{k})$, and $\psi(\mathbf{k})$ are parameters to be determined. We define $h_{D,m}(\mathbf{k})=[h_{4,m}(\mathbf{k})+h_{5,m}(\mathbf{k})]/2$, $h_{Q,m}(\mathbf{k})=[h_{4,m}(\mathbf{k})-h_{5,m}(\mathbf{k})]/2$, $\epsilon_\alpha(\mathbf{k})=\
\sqrt{[h_{1,m}(\mathbf{k})+h_{Q,m}(\mathbf{k})]^2+\alpha^2(\mathbf{k})}$, and $\epsilon_\beta(\mathbf{k})=\
\sqrt{[h_{1,m}(\mathbf{k})-h_{Q,m}(\mathbf{k})]^2+\beta^2(\mathbf{k})}$ and obtain the eigenvalues for $H_{mm}(\mathbf{k})$ to read
\begin{equation}
    \lambda_{1,2}(\mathbf{k})=h_{0,m}(\mathbf{k})+h_{D,m}(\mathbf{k})\pm \epsilon_\alpha(\mathbf{k})
\end{equation}
\begin{equation}
    \lambda_{3,4}(\mathbf{k})=h_{0,m}(\mathbf{k})-h_{D,m}(\mathbf{k})\pm \epsilon_\beta(\mathbf{k})
\end{equation}
The following two equations, determine $\alpha(\mathbf{k}),\ \phi(\mathbf{k})$ and $s(\mathbf{k})$:
\begin{equation}
    \begin{aligned}
        &\alpha(\mathbf{k})e^{i\phi(\mathbf{k})}=h_{2,m}(\mathbf{k})+h_{3,m}(\mathbf{k}),\\ &s(\mathbf{k})=\frac{h_{1,m}(\mathbf{k})+h_{Q,m}(\mathbf{k})+\epsilon_\alpha(\mathbf{k})}{\alpha(\mathbf{k})}
    \end{aligned}
\end{equation}
and the following two equations, determine $\beta(\mathbf{k}),\ \psi(\mathbf{k})$ and $t(\mathbf{k})$:
\begin{equation}
    \begin{aligned}
        &\beta(\mathbf{k})e^{i\psi(\mathbf{k})}=h_{2,m}(\mathbf{k})-h_{3,m}(\mathbf{k}),\\ &t(\mathbf{k})=\frac{h_{1,m}(\mathbf{k})-h_{Q,m}(\mathbf{k})+\epsilon_\beta(\mathbf{k})}{\beta(\mathbf{k})}
    \end{aligned}
\end{equation}
Evidently, the hyperspin degeneracy has been lifted by $\hat{H}$. More importantly, however, the $j_y=+\hbar/2$ hyperspin sector, parametrized by $s(\mathbf{k})$, $\phi(\mathbf{k})$, and $\alpha(\mathbf{k})$, is completely decoupled from the $j_y=-\hbar/2$ hyperspin sector, parametrized by $t(\mathbf{k})$, $\psi(\mathbf{k})$, and $\beta(\mathbf{k})$, whereby establishing the collinear bands in $\mathbf{k}$-space. The other two hyperspin components in the matrix form read
\begin{equation}
\hat{\mathcal{J}}_x(\theta)=\frac{\hbar}{2}
    \begin{pmatrix}
        0 & 1 & 0 & 0\\
        1 & 0 & 0 & 0\\
        0 & 0 & {\rm sin}\theta & {\rm cos}\theta\\
        0 & 0 & {\rm cos}\theta & -{\rm sin}\theta
    \end{pmatrix}
\end{equation}
\begin{equation}
\hat{\mathcal{J}}_z(\theta)=\frac{\hbar}{2}
    \begin{pmatrix}
        0 & 1 & 0 & 0\\
        1 & 0 & 0 & 0\\
        0 & 0 & {\rm cos}\theta & -{\rm sin}\theta\\
        0 & 0 & -{\rm sin}\theta & -{\rm cos}\theta
    \end{pmatrix}.
\end{equation}
Note that here angle $\theta$ between $\mathbf{M}_{\mathbb{A}}$ and $\mathbf{M}_{\mathbb{B}}$ is not a coordinate variable, but a physical parameter determined by magnetic properties of the HAM. It is straightforward to show that
\begin{equation}
    \langle\mathbf{v}_i|\mathcal{J}_x|\mathbf{v}_i\rangle=\langle\mathbf{v}_i|\mathcal{J}_z|\mathbf{v}_i\rangle=0,\ (i=1\ {\rm to}\ 4 )
\end{equation}
as expected. Additionally, one can show
\begin{equation}
    \begin{aligned}
        &\langle\mathbf{v}_1|\mathcal{J}_j|\mathbf{v}_2\rangle=\langle\mathbf{v}_2|\mathcal{J}_j|\mathbf{v}_1\rangle=0\\
        &\langle\mathbf{v}_3|\mathcal{J}_j|\mathbf{v}_4\rangle=\langle\mathbf{v}_4|\mathcal{J}_j|\mathbf{v}_3\rangle=0,\  (j=x,y,z).
    \end{aligned}
\end{equation}
For the physical spin, on the other hand, none of the analogous commutation relations holds for any component of the physical spin $\hat{\mathcal{S}}_j$, as detailed in Sec. V of Ref.~\cite{SM}. 

So far, our discussion considers only the diagonal $m=m'$ terms. When contributions from the off-diagonal $m\neq m'$ terms are included, first, the commutation relation $[\hat{H},\ {\hat{\mathcal{J}}}_y]=0$ still holds (see Sec. IV of Ref.~\cite{SM}); second, the eigen states become superpositions of momentum components within a fixed hyperspin sector, namely,
\begin{equation}
\label{psiu}
    |\Psi_{n\mathbf{k}}^\uparrow\rangle=\sum_{m}\Big[c^{n}_{1,m\mathbf{k}}|v_{1,m\mathbf{k}}(\mathbf{r})+c^{n}_{2,m\mathbf{k}}|v_{2,m\mathbf{k}}(\mathbf{r})\Big]
\end{equation}
\begin{equation}
\label{psid}
    |\Psi_{n\mathbf{k}}^\downarrow\rangle=\sum_{m}\Big[c^{n}_{3,m\mathbf{k}}|v_{3,m\mathbf{k}}(\mathbf{r})+c^{n}_{4,m\mathbf{k}}|v_{4,m\mathbf{k}}(\mathbf{r})\Big]
\end{equation}
where $c_{i,m\mathbf{k}}^n$ are the $\mathbf{k}$-space expansion coefficients, $|v_{i,m\mathbf{k}}(\mathbf{r})\rangle=\mathbf{v}_i\cdot|u_{\mathbb{A},m\mathbf{k}}^+(\mathbf{r})\rangle,|u_{\mathbb{A},m\mathbf{k}}^-(\mathbf{r})\rangle,|u_{\mathbb{B},m\mathbf{k}}^+(\mathbf{r})\rangle,|u_{\mathbb{B},m\mathbf{k}}^-(\mathbf{r})\rangle]T$, $\uparrow\downarrow$ here denote the two sectors of the hyperspin $\langle\hat{\mathcal{J}}_y\rangle$, which do not mix even when off-diagonal terms (see Sec. IV of Ref.~\cite{SM}) are included. Eqs.~(\ref{psiu}-\ref{psid}) guarantee that
\begin{equation}
    \begin{aligned}
        \langle\Psi_{n\mathbf{k}}^{\uparrow(\downarrow)}|\hat{\mathcal{S}}_{x(z)}|\Psi_{n\mathbf{k}}^{\uparrow(\downarrow)}\rangle&=0\\
        \mathcal{S}_{n,y}(\mathbf{k})=\langle\Psi_{n\mathbf{k}}^{\uparrow(\downarrow)}|\hat{\mathcal{S}}_{y}|\Psi_{n\mathbf{k}}^{\uparrow(\downarrow)}\rangle&\in[-\hbar/2,\ \hbar/2]
    \end{aligned}
\end{equation}
Here, although $\langle\hat{\mathcal{S}}_y\rangle$ is not quantized, $\langle\hat{\mathcal{S}}_x\rangle=\langle\hat{\mathcal{S}}_z\rangle=0$, irrespective of the fact that $\Lambda_m(\mathbf{k})\neq0$ (see Sec. V of Ref.~\cite{SM}).

\section[\label{sec.5}]{Discussion}
So far, only the simplest HAMs that contain two AFM subsystems are considered, see Fig.~\ref{fig2}. However, there is no reason to limit the number of subsystems to only two. Consider, for example, three AFM subsystems whose spin directions are along $\mathbf{M}_\mathbb{A}$, $\mathbf{M}_\mathbb{B}$, and $\mathbf{M}_\mathbb{C}$, respectively, see Fig.~\ref{fig1}(d). The question is whether a direction $\mathbf{y}$, assigned to ${\hat{\mathcal{J}}}_y$ for the combined system, exists. If we first pick $\mathbf{M}_\mathbb{A}$ and $\mathbf{M}_\mathbb{B}$ and define $\mathbf{y}=\mathbf{M}_\mathbb{A}\times\mathbf{M}_\mathbb{B}$, then when including $\mathbf{M}_\mathbb{C}$, it requires that $\mathbf{y}\bot\mathbf{M}_\mathbb{C}$. We can then cycle $\mathbb{A}, \mathbb{B}$, and $\mathbb{C}$ to find $\mathbf{y}\bot\mathbf{M}_\mathbb{B}$ and $\mathbf{y}\bot\mathbf{M}_\mathbb{A}$. In 3-dimensional space, if $\mathbf{M}_\mathbb{A}$, $\mathbf{M}_\mathbb{B}$, and $\mathbf{M}_\mathbb{C}$ are independent vectors, this is impossible. However, if $\mathbf{M}_\mathbb{A}$, $\mathbf{M}_\mathbb{B}$, and $\mathbf{M}_\mathbb{C}$ are coplanar, a common $\mathbf{y}$ would be possible and, under the $\Theta\bm{\tau}$ symmetry, the commutation relation $[\hat{H},\ {\hat{\mathcal{J}}}_y]$ would hold. 

Note that in all discussions, we do not require magnetic atoms in different subsystems to be the same, nor their magnitudes $|\mathbf{M}_{\mu}|$. If, however, we do have symmetry that relates magnetic atoms in different subsystems and provides that all the local magnetic moments are coplanar, we then would have more odd-wave magnetic states such as f-wave, h-wave, etc., rather than the simplest p-wave state.
In essence, the onset of a collinear spin splitting in $\mathbf{k}$-space is a manifestation of the existence of a conserved spin-momentum-related quantity in the system. For AMs, this quantity is the physical spin, say $\hat{\mathcal{S}}_y$, in addition to a space rotation or mirror symmetry. In contrast, for HAMs, this quantity is the hyperspin $\hat{\mathcal{J}}_y$, which is defined when and only when all the local magnetic moments are coplanar.

Recently, it was proposed \cite{song2025unified} that a spin-space-group $\{\mathcal{C}_{2y}|\bm{\tau}\}$ symmetry, where $\mathcal{C}_{2y}$ is a spin rotation by angle $\pi$ about the $y$ axis, which bisects $\mathbf{M}_{\rm A}$ and $\mathbf{M}_{\rm B}$, could also yield a collinear spin splitting in $\mathbf{k}$-space in non-collinear magnets. But this collinearity only exists within the $k_y=0$ plane and is not alternating. Nevertheless, it is legitimate to ask whether a symmetry other than $\Theta\bm{\tau}$ can also give rise to alternated collinear spin splitting. Because collinear spin splitting refers to energies at the same $\mathbf{k}$ point, a spin-space-group symmetry $\{R_\beta|R_i\}$ with any spin part $R_\beta$ and a nontrivial spatial part $R_i$ that transforms $\mathbf{k}$ to a different point in Brillouin zone, can be excluded. Further, $\{R_\beta|\bm{\tau}\}$ with a trivial spatial part cannot lead to alternate spin splitting, hence it can also be excluded. Therefore, as far as we can tell, $\Theta\bm{\tau}$ is unique for the emergence of HAMs.

To conclude, we build a spin Hamiltonian from first principles for hyperspin altermagnets. We show why $\Theta\bm{\tau}$ leads to spin degeneracy for conventional antiferromagnets, but not for hyperspin altermagnets, including p-wave hyperspin altermagnets \cite{hellenes2023p}. We propose a hyperspin and analytically prove that the emergence of collinear spin splitting in $\mathbf{k}$-space is a result of the commutation relation between hyperspin and Hamiltonian, and the reason for the lack of spin quantization is due to momentum-hyperspin wavefunction entanglement. The logic and approach developed here can also be generalized to other magnets with more complex magnetic structures. Based on the theory, a brief $\Theta\bm{\tau}$ symmetry analysis of existing materials reveals that CeNiAsO \cite{hellenes2023p,zhou2025anisotropic}, the U$_2$Ni$_2$In family of alloys, including Yb$_2$Pd$_2$(In$_{0.4}$Sn$_{0.6}$) and Tb$_2$Pd$_{2.05}$Sn$_{0.95}$, CuSb$_2$O$_6$, CeCo$_2$Ge$_4$O$_{12}$, ErAuIn/TbAuIn, and LaMnAu$_5$ are promising candidates for hyperspin altermagnets.

\begin{acknowledgments}
We thank Dr. Yongzhen Xu for helpful discussions. H.-Y. Ma acknowledges the financial support from National Natural Science Foundation of China (Grant No. 12504225), Guangdong Provincial Quantum Science Strategic Initiative (GDZX2401001). J.-F. J. thanks the NSFC under Grants No. 12488101 for financial support.
\end{acknowledgments}


\begin{thebibliography}{33}
\bibitem [{not()}]{note1}%
  \BibitemOpen
  \href@noop {} {}\bibinfo {note} {For clarity, we label the quantization
  direction as the y direction}\BibitemShut {NoStop}%
\end{thebibliography}
%

\end{document}